\documentclass[]{article}    %use this line if you want to compile in LaTeX2e
\usepackage{graphicx}        %use for LaTeX2e
\def\title#1{\centerline{\Large \bf #1}\kern10.pt}
\def\subtitle#1{\centerline{\Large \bf #1}\kern10.pt}
\def\author#1{\centerline{#1}}
\def\address#1{\centerline{\em #1}}

\begin{document}
%\stamp
\small
%
% ZZZZZZZZZZZZZ
% This section can be modified to include text, figures and tables.
%
\title{Exploring the BEC-BCS Crossover}
\vspace{-0.1in} \subtitle{with an Ultracold Gas of $^6$Li Atoms}
\author{M. Bartenstein,$^1$ A. Altmeyer,$^1$ S. Riedl,$^1$ S. Jochim,$^1$ R. Geursen,$^1$}
\author{C. Chin,$^1$ J. Hecker Denschlag,$^1$ and R. Grimm$^{1,\,2}$}
\address{$^1$Institute of Experimental Physics, Innsbruck University, Innsbruck, Austria}
\address{$^2$Institute for Quantum Optics and Quantum Information (IQOQI),}
\address{Austrian Academy of Sciences, Innsbruck, Austria}
\vskip10.pt \centerline{\bf Abstract} We present an overview of
our recent measurements on the crossover from a Bose-Einstein
condensate of molecules to a Bardeen-Cooper-Schrieffer superfluid.
The experiments are performed on a two-component spin-mixture of
$^6$Li atoms, where a Fesh\-bach resonance serves as the
experimental key to tune the s-wave scattering length and thus to
explore the various interaction regimes. In the BEC-BCS crossover,
we have characterized the interaction energy by measuring the size
of the trapped gas, we have studied collective excitation modes,
and we have observed the pairing gap. Our observations provide
strong evidence for superfluidity in the strongly interacting
Fermi gas.

\section{Introduction}

The crossover from a Bose-Einstein condensate (BEC) to a
Bardeen-Cooper-Schrieffer (BCS) superfluid has for decades attracted
considerable attention in condensed matter theory
\cite{eagles,leggett,NSR,levin}. The observation of BEC of molecules in
ultracold trapped Fermi gases of $^6$Li and $^{40}$K \cite{li2becinn, k2bec,
li2becMIT, li2becENS, huletBEC} has opened up a unique route to explore this
BEC-BCS crossover. Magnetically tuned scattering resonances, known as Feshbach
resonances \cite{feshbach}, play the key role to control the two-body
interaction and to vary the coupling strength over a very broad range.
Exploiting Feshbach tuning, recent experiments have begun to explore the
crossover by studying elementary properties of the system under variable
interaction conditions. The internal interaction energy was measured by
detecting the cloud size of a trapped gas \cite{markus1} and by observing the
expansion of the gas after release \cite{li2becENS}. The condensed nature of
fermionic atom pairs was demonstrated by rapid conversion of the ``Fermi
condensate'' into a molecular BEC \cite{jin2004,mit2004}. The study of
collective excitation modes \cite{kinast2004,markus2,kinast2004b} provided
first insight into changes of the equation of state and hydrodynamics of the
system in the crossover. Spectroscopic measurements of the pairing energy
\cite{chin2004} showed the crossover from molecular pairing to fermionic
``Cooper'' pairing. The results of these experiments provide strong evidence of
``resonance superfluidity'' \cite{holland,timmermans,ohashi,stajic} in a
strongly interacting Fermi gases.

Here we summarize the BEC-BCS crossover experiments that we recently performed
in Innsbruck on an ultracold gas of $^6$Li atoms. Degenerate Fermi gases of
$^6$Li have been produced by several groups
\cite{DFGRice,DFGENS,DFGDuke,DFGMIT,li2becinn}. A mixture of the two lowest
spin states of $^6$Li is particularly interesting because it is stable against
two-body decay and it exhibits a broad Feshbach resonance \cite{houbiers} in
combination with a narrow one \cite{huletnarrow}. The broad resonance has been
widely employed by us and other groups to create a strongly interacting Fermi
gas \cite{thomasscience}, to study interaction effects \cite{bourdel}, to form
stable weakly bound molecules \cite{ENSmolecules,IBKmolecules}, to create
Bose-Einstein condensates of molecules
\cite{li2becinn,li2becMIT,li2becENS,huletBEC}, and to study the BEC-BCS
crossover \cite{markus1,mit2004,markus2,kinast2004b,chin2004}. After obtaining
the experimental data summarized here, we recently performed high-precision
spectroscopic experiments on the $^6$Li molecular interaction parameters
\cite{markus3}. We here discuss our experimental crossover data
\cite{markus1,markus2,chin2004} with up-to-date knowledge of the two-body
scattering properties.

\section{Creation of a molecular BEC}

Our molecules are weakly bound $^6$Li$_2$ dimers in the last bound level very
close to the dissociation threshold. A key property of these molecules is their
stability against inelastic decay resulting from the fermionic nature of the
constituent atoms \cite{petrov}. %The molecular binding energy $E_{\rm b}$ is related to the
%s-wave scattering length $a$ by $E_{\rm b} = $.
For molecule formation and evaporative cooling towards BEC, we choose a
magnetic field of 764\,G, which is below the center of the resonance at 834\,G
\cite{markus3}. Here the s-wave scattering length is $a=4500\,a_0$ and the
molecular binding energy is $E_{\rm b}= \hbar^2/(ma^2) = k_{\rm B}\times
1.4\,\mu K$, where $a_0$ denotes Bohr's radius and $k_{\rm B}$ is Boltzmann's
constant.

Evaporative cooling is performed in a single-beam optical dipole trap
\cite{li2becinn}. A 1030-nm laser beam, which is focused to a waist of
25$\mu$m, is loaded with $2 \times 10^6$ optically precooled atoms. For forced
evaporative cooling, the laser power is exponentially ramped down from the
initial value of 10\,W to a few mW with a time constant of 460\,ms. The
molecules are formed during the evaporation following a chemical atom-molecule
equilibrium \cite{chin03,french}. Finally, all remaining atoms form molecules
and the pure molecular sample condenses into a BEC. As a consequence of
molecular Bose-Einstein condensation we first observed that, at the end of the
evaporation, the number of atoms confined in a very shallow trap exceeded the
number of quantum states available for fermionic atoms by almost an order of
magnitude \cite{li2becinn}. We furthermore observed a collective excitation
mode and demonstrated magnetically tuned mean-field effects \cite{li2becinn}.
Using in-situ absorption imaging we also observed the phase transition in
characteristic bimodal distributions of the trapped cloud \cite{markus1}; see
Fig.~1. The images provide a lower bound for the condensate fraction of 90\%
and thus demonstrate that an essentially pure molecular BEC is formed.

\begin{figure}[h]
\begin{center}
\includegraphics[width=3in]{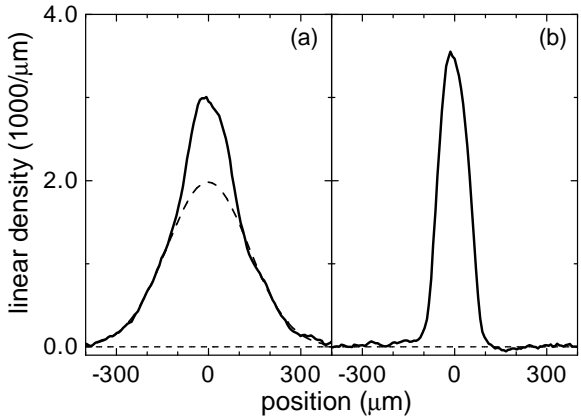}
\end{center}
\vspace{-0.2in} \caption{\small Molecular BEC is observed using
{\it in situ} axial density profiles of the trapped cloud
\cite{markus1}. (a) A partially condensed cloud with $4 \times
10^5$ molecules is obtained when the evaporation is stopped at a
final laser power of 28\,mW, and (b) an essentially pure BEC of
$2\times 10^5$ molecules is obtained at a final laser power of
3.8\,mW.}
\end{figure}

We measured very long lifetimes of the molecular BEC. After recompression in
the optical dipole trap to a peak number density of $1.0(5) \times 10^{13} {\rm
cm}^{-3}$ we observed a lifetime of $40$s. This corresponds to a very low upper
bound for the binary loss coefficient of $1\times 10^{-14}$\,cm$^3$/s, which
demonstrates the enormous stability of weakly bound $^6$Li$_2$ molecules
\cite{ENSmolecules,IBKmolecules,petrov}.

The pure molecular BEC has served as a starting point for all our experiments
on the BEC-BCS crossover, as described in the following.

\section{\bf Conversion of a molecular BEC to a degenerate Fermi gas}

For exploring the crossover to a Fermi gas we apply slow magnetic
field ramps. We typically change the magnetic field from the BEC
production value to a final value in 1\,s. This is slow enough for
the gas to react adiabatically. In a test experiment
\cite{markus1} we checked the reversibility of the crossover
process by linearly ramping up the magnetic field from 764\,G to
1176\,G and down again to 764\,G within 2\,s. The comparison of
the spatial profiles did not show any significant deviations. This
experiment showed that the conversion into a degenerate Fermi gas
and its back-conversion into a molecular BEC takes place without
loss and heating in a fully reversible way. This remarkable
possibility to change the character of the gas in a reversible and
isentropic way highlights the outstanding properties of $^6$Li
Fermi gases for BEC-BCS crossover studies. Note that the
isentropic conversion of a molecular BEC into a degenerate Fermi
gas goes along with a substantial temperature reduction
\cite{carr}.

In a first set of experiments we measured spatial profiles of the
trapped ultracold gas for magnetic fields between 740\,G and
1440\,G \cite{markus1}. These profiles provide information on the
interaction energy in the sample and on the equation of state. The
recorded axial density profiles are in general well fit by Thomas
Fermi profiles. Fig.~3(b) shows how the measured root-mean-square
axial size $z_{\rm rms}$ changes with the magnetic field. For
comparison, Fig.~3(a) displays the magnetic-field dependence of
the atomic scattering length $a$ \cite{markus3}. Up to 950\,G the
observed increase in $z_{\rm rms}$ is due to the crossover from
the molecular BEC to the degenerate Fermi gas. For higher magnetic
fields, the shrinking of the axial cloud size is caused by an
increasing magnetic confinement \cite{markus1}. To remove the
explicit trap dependence, we normalize the observed size to the
one expected for a non-interacting Fermi gas. In Fig.~3(c) we show
the normalized axial size $\zeta=z_{\rm rms}/z_0$, where $z_0$ is
the rms axial size of a non-interacting zero-temperature Fermi gas
with $N = 4\times10^5$ atoms.

\begin{figure}[h]
\begin{center}
\includegraphics[width=2.5in]{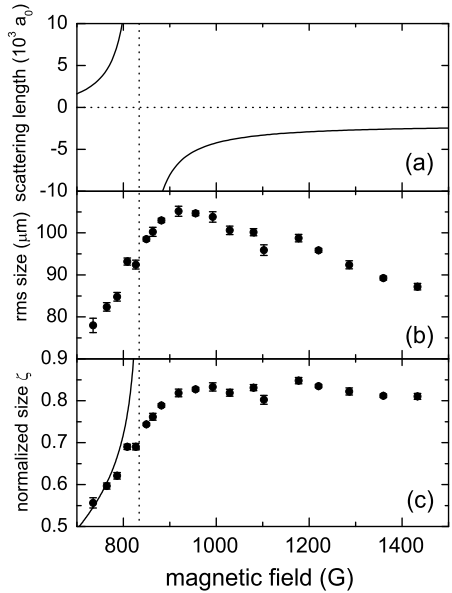}
\end{center}
\vspace{-0.2in} \caption{\small Cloud size measurements across the Feshbach
resonance \cite{markus1}. In (a) the atomic scattering length $a$ is shown
according to \cite{markus3}; the resonance at 834\,G is marked by the vertical
dashed line. The data in (b) display the measured rms cloud sizes. In (c), the
same data are plotted after normalization to a non-interacting Fermi gas. The
solid line shows the expectation from BEC mean-field theory with $a_{\rm mol} =
0.6\,a$.}
\end{figure}

Below the Feshbach resonance, the observed dependence of the cloud size agrees
well with the mean-field behavior of a molecular BEC in the Thomas-Fermi limit.
In this regime, the normalized size is given by $\zeta=0.688 (a_{\rm
mol}/a)^{1/5} (E_{\rm F}/E_{\rm b})^{1/10}$, where $E_{\rm F}$ is the Fermi
energy of the gas without interactions. Fig.~3(c) shows the corresponding curve
(solid line) calculated with a molecule-molecule scattering length $a_{\rm
mol}/a=0.6$ \cite{petrov}. We find that this BEC limit provides a reasonable
approximation up to $\sim$780\,G; here the molecular gas interaction parameter
is $n_{\rm mol}a_{\rm mol}^3 \approx 0.05$. Alternatively, the interaction
strength can be expressed as $k_Fa\approx 2$, where the Fermi wavenumber
$k_{\rm F}$ is related to the Fermi energy by $E_{\rm F} = \hbar^2 k_{\rm
F}^2/(2m)$. The crossover to the Fermi gas is observed in the vicinity of the
Feshbach resonance between 780\,G and 950\,G; here $\zeta$ smoothly increases
with the magnetic field until it levels off at 950\,G, where the interaction
strength is characterized by $k_{\rm F}a \approx -1.4$. Our results show that
the crossover occurs within the range of $-0.7 < (k_{\rm F}a)^{-1}< 0.7$, which
corresponds to the strongly interacting regime.

The case of resonant two-body interaction is of particular
interest for many-body quantum physics. For $|a|\rightarrow
\infty$ a universal regime is realized
\cite{heiselberg,universality,thomasmechan}, where scattering is
fully governed by unitarity and the scattering length drops out of
the description. The normalized cloud size can be written as
$\zeta=(1+\beta)^{1/4}$, where $\beta$ parameterizes the
mean-field contribution to the chemical potential in terms of the
local Fermi energy \cite{thomasmechan}. At 834\,G, our
measurements show $\zeta=0.72\pm0.07$ which provides
$\beta=-0.73^{+0.12}_{-0.09}$. Here the total error range includes
our best knowledge of statistic and systematic uncertainties, with
the particle number giving the dominant contribution to the error
budget. Recent quantum Monte Carlo calculations yielded $\beta =
-0.56(1)$ \cite{carlson} and $-0.58(1)$ \cite{astrakharchik}. Our
experimental results are close to these predictions, but they show
a deviation somewhat larger than our assumed experimental error
range. If we, in turn, use the theoretical predictions on $\beta$
to calibrate our particle number we obtain $2.0\times10^5$ trapped
atoms, which only slightly falls outside of our estimated
uncertainty range.

It is very interesting to compare our data with recent predictions from
advanced crossover theories. Ref.~\cite{strinatiprofiles} compares spatial
profiles calculated from a diagrammatic theory including pairing fluctuations
beyond mean-field with our experimental profiles. The diagrammatic theory
itself \cite{strinaticompare} agrees very well with quantum Monte Carlo
calculations \cite{astrakharchik}, in particular in the range of our
experiments. If we assume $2\times10^5$ atoms for our experiments, our data on
spatial profiles and the size of the trapped gas are found to agree excellently
with theory in the whole crossover range.

\section{\bf Measurements of collective oscillations}

The investigation of collective excitation modes \cite{stringari96} is well
established as a powerful method to gain insight into the physical behavior of
ultracold quantum gases in different regimes of Bose \cite{stringaribook} and
Fermi gases \cite{excite}. Ref.~\cite{stringari04} pointed out an interesting
dependence of the collective frequencies in the BEC-BCS crossover of a
superfluid Fermi gas. Superfluidity implies a hydrodynamic behavior which can
cause substantial changes in the excitation spectrum and in general very low
damping rates. However, in the crossover regime the strong interaction between
the particles also results in hydrodynamic behavior in the normal,
non-superfluid phase. Therefore the interpretation of collective modes in the
BEC-BCS crossover in terms of superfluidity is not straightforward and needs
careful investigation to identify the different regimes.

We studied two elementary collective excitation modes in the
BEC-BCS crossover \cite{markus2}. The slow {\em axial} compression
mode confirmed theoretical expectations on the equation of state
of the gas in the crossover and is in general well understood
\cite{tosi04}. In the BEC limit we observed an oscillation
frequency $\Omega_z$ consistent with the expected $\Omega_z =
\sqrt{5/2} \omega_z = 1.581 \omega_z$, where $\omega_z$ denotes
the axial trap frequency. In the unitarity limit (at 834\,G) we
measured a 2\% down-shift of the axial frequency, in agreement
with the prediction $\Omega_z = \sqrt{5/2} \omega_z = 1.549
\omega_z$. This down-shift is a consequence of the fact that the
equation of state of the quantum gas changes from $\mu \propto n$
to $\mu \propto n^{2/3}$ \cite{stringari04,tosi04}. With the
magnetic field increasing beyond the resonance, we observe a
further decrease in the collective excitation frequency until a
minimum is reached at about 900\,G, where $1/(k_Fa) \approx -0.5$.
This observation is in agreement with theoretical expectations
\cite{tosi04}. With further increasing magnetic field and
decreasing interaction strength, we observe a gradual transition
to a collisionless regime and a corresponding gradual increase of
the collective frequency toward $\Omega_z/\omega_z=2$.

At about 815\,G, somewhat below the resonance point, we observed
extremely low damping rates for the axial collective oscillations.
Here the $1/e$ damping time corresponded to about 160
oscillations. This striking behavior may be seen as a strong piece
of evidence for {\it superfluidity}.

\begin{figure}[hhhh]
\begin{center}
\includegraphics[width=3in]{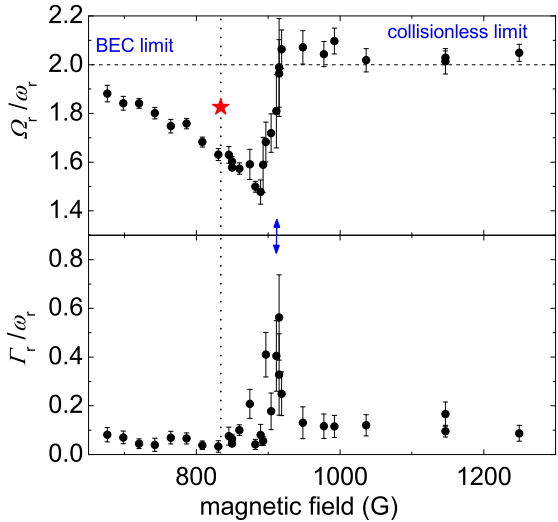}
\end{center}
\vspace{-0.2in} \caption{\small Measured frequency $\Omega_r$ and damping rate
$\Gamma_r$ of the radial compression mode, normalized to the trap frequency
(sloshing mode frequency) $\omega_r$. In the upper graph, the dashed line
indicates $\Omega_r/\omega_r=2$, which corresponds to both the BEC limit and
the collisionless Fermi gas limit. The vertical dotted line marks the resonance
position at $834$\,G. The star indicates the theoretical expectation of
$\Omega_r/\omega_r=\sqrt{10/3}$ in the unitarity limit. A striking change in
the excitation frequency occurs at $\sim$910\,G (arrow) and is accompanied by
anomalously strong damping.}
\end{figure}

The fast {\em radial} compression mode showed a surprising dependence on the
magnetic field \cite{markus2}, which is not fully understood yet. Fig.~3 shows
the observed radial oscillation frequency $\Omega_r$ and damping rate
$\Gamma_r$, normalized to the radial trap frequency $\omega_r$. The unitarity
gas shows a large down-shift to $\Omega_r/\omega_r = 1.67(3)$, much larger than
the expected value of $\Omega_r/\omega_r = \sqrt{10/3} = 1.826$ from
hydrodynamic theory \cite{stringari04}. However, it is questionable whether
simple hydrodynamic theory still applies for our experimental conditions
\cite{combescotcomment}.

The most striking feature of our radial oscillation data is the
abrupt change of the excitation frequency at a field of
$\sim$910\,G, which is accompanied by an anomalously large
damping. Beyond this point the collective oscillations exhibit the
collisionsless value of $\Omega_r/\omega_r = 2$, which clearly
shows that the gas loses its hydrodynamic properties. A plausible
interpretation \cite{combescotcomment} of this breakdown of
hydrodynamics is a coupling of the collective oscillations to the
fermionic pairs in the strongly interacting gas. This
interpretation is supported by our measurements on pairing
energies \cite{chin2004}, see below.

The Duke group has performed experiments on radial collective
oscillations in a set-up very similar to ours
\cite{kinast2004,kinast2004b}. They found a strong temperature
dependence of the damping rate, which provides further evidence
for superfluidity in the system \cite{kinast2004}. On resonance,
they measured a collective frequency in agreement with
hydrodynamic theory and thus not consistent with our observation.
Beyond the Feshbach resonance the Duke measurements confirmed our
observation of a breakdown of hydrodynamics \cite{kinast2004b}.
Clearly more measurements are needed on the radial excitation
mode.

\section{\bf Observation of the pairing gap}

The formation of pairs generally represents a key ingredient of superfluidity
in fermionic systems. The ``pairing gap'', which corresponds to the binding
energy of the pairs, is a central quantity to characterize the pairing regime.
We employed radio-frequency (rf) spectroscopy
\cite{debbieRF,ketterleRF,jinnature1} to study pairing in the BEC-BCS
crossover. Spectral signatures of pairing were theoretically considered in
Refs.\ \cite{zollergap1,paivi1,buechler,honew,paivi2}. A clear signature of the
pairing process is the emergence of a double-peak structure in the spectral
response as a result of the coexistence of unpaired and paired atoms. The
pair-related peak is located at a higher frequency than the unpaired-atoms
signal as energy is required for pair breaking.

We performed rf spectroscopy by driving transitions of the nuclear
spin to an empty state \cite{chin2004}. The loss of atoms from the
two-component spin-mixture as a function of the radio frequency
constitutes our spectroscopic signal. We recorded rf spectra for
different degrees of cooling and in various coupling regimes
(Fig.~4). We realize the molecular regime at $B=720$\,G ($a =
+2200\,a_0$). For the resonance region, we examined two different
magnetic fields, $B=822$\,G ($a \approx +35,000\,a_0$) and
$B=837$\,G ($a \approx -100,000\,a_0$). We also studied the regime
beyond the resonance with large negative scattering length at
$B=875$\,G ($a = -1200\,a_0$).

At a ``high'' temperature $T\approx 6T_{\rm F}$ ($T_{\rm F} =
15\mu$K) we just observe the narrow atomic transition line (upper
row in Fig.~4) without any effect of interactions. This line
serves us as a frequency reference, and we present our spectra as
a function of the rf offset with respect to this atomic frequency.

Already at $T'/T_{\rm F} = 0.5$ \cite{temperature}, we observed
the double-peak structure characteristic for the onset of pairing
(middle row in Fig.~4). In the molecular regime ($B=720$\,G), the
sharp atomic peak is well separated from the broad dissociation
signal \cite{chinjulienne}, which shows a molecular binding energy
of $E_{\rm b} = h \times 130\,{\rm kHz} = k_{\rm B} \times
6.2\,\mu$K. For $B$ approaching the resonance position, the peaks
begin to overlap. In the resonance region (822 and 837\,G), we
still observe a relatively narrow atomic peak at the original
position together with a pair signal. For magnetic fields beyond
the resonance, we can resolve the double-peak structure for fields
up to $\sim$900\,G.

\begin{figure}[t]
\begin{center}
\includegraphics[width=4.5in]{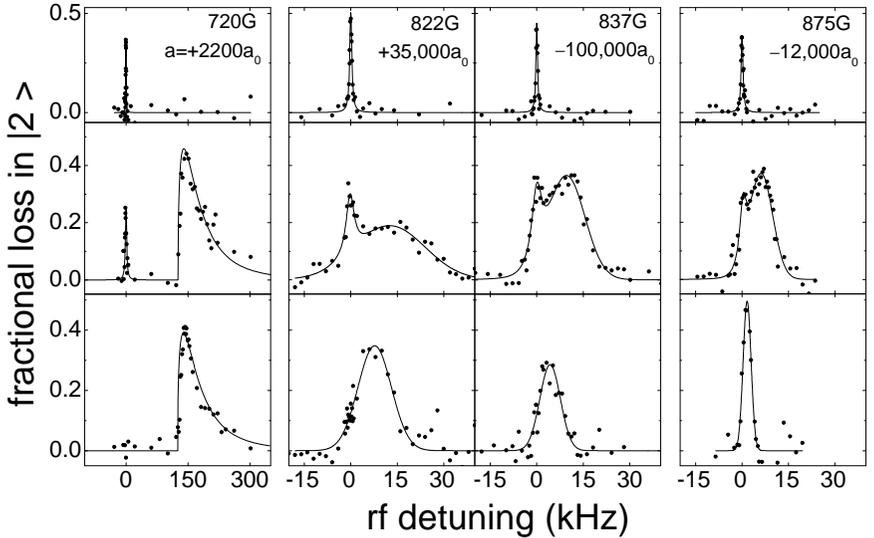}
\end{center}
\vspace{-0.2in} \caption{\small Radio-frequency spectra for
various magnetic fields and different degrees of evaporative
cooling. The rf offset is given relative to the atomic transition.
The molecular limit is realized for $B=720$\,G (first column). The
resonance regime is studied for $B=822$\,G and 837\,G (second and
third column). The data at 875\,G (fourth column) explore the
crossover on the BCS side. Upper row, signals of unpaired atoms at
$T'\approx 6 T_{\rm F}$ ($T_{\rm F} =15\,\mu$K); middle row,
signals for a mixture of unpaired and paired atoms at $T' =
0.5T_{\rm F}$ ($T_{\rm F} =3.4\,\mu$K); lower row, signals for
paired atoms at $T' < 0.2T_{\rm F}$ ($T_{\rm F} =1.2\,\mu$K). The
solid lines are introduced to guide the eye.}
\end{figure}
%\vspace{-0.25in} \noindent Figure 4: Axial density profiles of a partially

At a very low temperature $T'/T_{\rm F} < 0.2$ \cite{temperature},
realized with deep evaporative cooling, we observed a
disappearance of the narrow atomic peak in the rf spectra (lower
row in Fig.~4). This shows that essentially all atoms are paired.
In the BEC limit ($720\,$G) the dissociation lineshape is
identical to the one observed in the trap at higher temperature
and Fermi energy. Here the localized pairs are molecules with a
size much smaller than the mean interparticle spacing, and the
dissociation signal is independent of the density. In the
resonance region (822 and 837\,G) the pairing signal shows a clear
dependence on density (Fermi energy), which becomes even more
pronounced beyond the resonance (875\,G).

For understanding the spectra in the fermionic many-body regime on resonance
and beyond both the homogeneous lineshape of the pair signal
\cite{paivi1,honew} and the inhomogeneous line broadening due to the density
distribution in the harmonic trap need to be taken into account \cite{paivi2}.
As an effect of inhomogeneity, fermionic pairing due to many-body effects takes
place predominantly in the central high-density region of the trap, and
unpaired atoms mostly populate the outer region of the trap where the density
is low \cite{bulgac,strinati,paivi2}. The spectral component corresponding to
the pairs thus shows a large inhomogeneous broadening in addition to the
homogeneous width of the pair-breaking signal. For the unpaired atoms the
homogeneous line is narrow and the effects of inhomogeneity and mean-field
shifts are negligible. These arguments explain why the RF spectra in general
show a relatively sharp peak for the unpaired atoms together with a broader
peak attributed to the pairs.

To quantitatively investigate the crossover from the two-body
molecular regime to the fermionic many-body regime we measure the
pairing energy in a range between 720\,G and 905\,G. The
measurements were performed after deep evaporative cooling
($T'/T_{\rm F} < 0.2$) for two different Fermi temperatures
$T_{\rm F}=1.2\,\mu$K and $3.6\,\mu$K (Fig.~5). As an effective
pairing gap we define $\Delta\nu$ as the frequency difference
between the pair-signal maximum and the bare atomic resonance. In
the BEC limit, the effective pairing gap $\Delta\nu$ simply
reflects the molecular binding energy $E_{\rm b}$ (solid line in
Fig.~5). With increasing magnetic field, in the BEC-BCS crossover,
$\Delta\nu$ shows an increasing deviation from this low-density
molecular limit and smoothly evolves into a density-dependent
many-body regime where $h\Delta\nu < E_{\rm F}$.

\begin{figure}[hhhh]
\begin{center}
\includegraphics[width=3in]{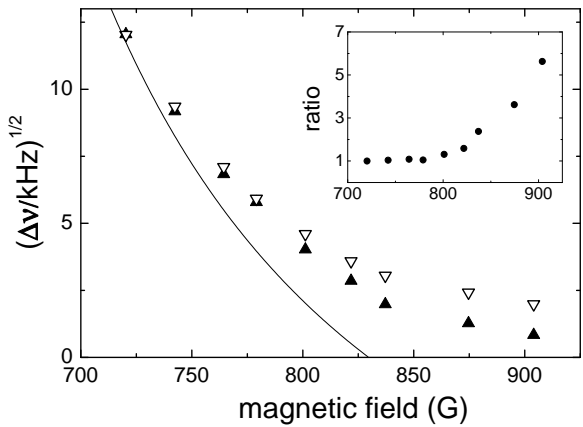}
\end{center}
\vspace{-0.2in} \caption{\small Measurements of the effective
pairing gap $\Delta\nu$ as a function of the magnetic field $B$
for deep evaporative cooling and two different Fermi temperatures
$T_{\rm F}=1.2\mu$K (filled symbols) and $3.6\mu$K (open symbols).
The solid line shows $\Delta\nu$ for the low-density limit where
it is essentially given by the molecular binding energy. The inset
displays the ratio of the effective pairing gaps measured at the
two different Fermi energies.}
\end{figure}
%\vspace{-0.25in} \noindent Figure 5: Axial density profiles of a partially

A comparison of the pairing energies at the two different Fermi
energies (inset in Fig.~5) provides further insight into the
nature of the pairs. In the BEC limit, $\Delta\nu$ is solely
determined by $E_{\rm b}$ and thus does not depend on $E_{\rm F}$.
In the universal regime on resonance, $E_{\rm F}$ is the only
energy scale and we indeed observe the effective pairing gap
$\Delta\nu$ to increase linearly with the Fermi energy. We find a
corresponding relation $h\Delta\nu \approx 0.2\,E_{\rm F}$. Beyond
the resonance, where the system is expected to change from a
resonant to a BCS-type behavior, $\Delta\nu$ is found to depend
more strongly on the Fermi energy and the observed gap ratio
further increases. We interpret this in terms of the increasing
BCS character of pairing, for which an exponential dependence
$h\Delta\nu/E_{\rm F} \propto \exp(-\pi/2k_F|a|)$ is expected.

Our measurements on the effective pairing gap $\Delta\nu$ also support a
possible interpretation of the abrupt frequency change of radial collective
oscillations (see preceding section). When the magnetic field is increased to
910\,G, we find the decreasing $\Delta\nu$ to reach the frequency of the radial
compression mode ($\sim$1\,kHz for the filled triangles in Fig.~5 and the
conditions of Fig.~3). We conclude that the oscillations can then couple to the
fermionic pairs, which leads to pair breaking and heating. This interpretation
is further supported by measurements at higher Fermi energies (open triangles
in Fig.~5 and collective mode measurements in Ref.~\cite{markusdr}), which
again show the abrupt change when the collective oscillation frequency meets
the effective pairing gap.

Radio-frequency spectroscopy in a harmonically trapped Fermi gas
with resonant interactions was theoretically analyzed in
Ref.~\cite{paivi2}. The calculated RF spectra demonstrate how a
double-peak structure emerges as the gas is cooled below $T/T_{\rm
F}\approx0.5$ and how the atomic peak disappears with further
decreasing temperature. In particular, the work addresses the role
of the ``pseudo-gap'' regime \cite{levin,stajic}, in which pairs
are formed before superfluidity is reached. According to the
calculated spectra, the atomic peak disappears at temperatures
well below the critical temperature for the phase-transition to a
superfluid. A recent theoretical study of the BCS-BEC crossover at
finite temperature \cite{strinati} predicts the phase-transition
to a superfluid to occur at a temperature that on resonance is
only $\sim$30\% below the point where pair formation sets in. Our
rf spectra thus provide strong evidence for superfluidity in the
strongly interacting Fermi gas.

\section{\bf Conclusion}

We have studied elementary macroscopic and microscopic properties
of an ultracold gas in the BEC-BCS crossover. For resonant
two-body interactions, we have obtained strong evidence for
superfluidity by measuring low damping rates of collective
oscillations and by observing the onset of pairing early in the
evaporative cooling process. This, together with complementary
observations made by other groups
\cite{jin2004,mit2004,kinast2004,li2becENS}, opens up intriguing
prospects for further experiments on the fascinating properties of
fermionic quantum matter.

\pagebreak
%\bigskip
%\medskip
\noindent
{\bf \normalsize Acknowledgements}

\bigskip \noindent We acknowledge support by the Austrian Science Fund (FWF)
within SFB 15 (project part 15) and by the European Union in the frame of the
Cold Molecules TMR Network under Contract No.\ HPRN-CT-2002-00290. C.C.\ is a
Lise-Meitner research fellow of the FWF.


\bigskip
\begin{thebibliography}{29}

%\bibitem{c1}
%A. Einstein, Phys. Rev. A {\bf 00}, 0101 (1911).

%\bibitem{c2} S. Timoshenko and S. Woinowsky-Krieger, {\it Theory of
%Plates and Shells}, McGraw-Hill, New York (1959).



%%%%% BEC-BCS crossover

\bibitem{eagles}
D. M. Eagles, Phys. Rev. {\bf 186}, 456 (1969).

\bibitem{leggett}
A. J. Leggett, in {\it Modern Trends in the Theory of Condensed Matter}, edited
by A. Pekalski and R. Przystawa, Springer-Verlag, Berlin (1980), pp. 13-27.

\bibitem{NSR}P. Nozi\`eres and S. Schmitt-Rink, {J. Low Temp. Phys.} {\bf 59}, 195 (1985).

\bibitem{levin}
Q. Chen, J. Stajic, S. Tan and K. Levin, cond-mat/0404274.

%\bibitem{Cho}
%See News Focus article by A. Cho, {\it Science\/} {\bf 301}, 750 (2003).




%%%%%%%%% molecular BEC %%%%%%%%%%%%%%%%%
\bibitem{li2becinn}
S. Jochim {\it et al.}, %M. Bartenstein, A. Altmeyer, G. Hendl, S. Riedl, C. Chin, J. Hecker Denschlag and R. Grimm,
Science {\bf 302}, 2101 (2003); published online 13 Nov 2003
(10.1126/science.1093280).

\bibitem{k2bec}
M. Greiner, C. A. Regal, and D. S. Jin, Nature {\bf 426}, 537 (2003).

\bibitem{li2becMIT}
M. W. Zwierlein {\it et al.}, %, C. A. Stan, C. H. Schunck, S. M. F. Raupach, S. Gupta, Z. Hadzibabic,
%and W. Ketterle,
Phys. Rev. Lett. {\bf 91}, 250401 (2003).

\bibitem{li2becENS}
T. Bourdel {\it et al.}, %L. Khaykovich, J. Cubizolles, J. Zhang, F. Chevy, M. Teichmann, L.
%Tarruell, S.J.J.M.F. Kokkelmans, and C. Salomon,
Phys. Rev. Lett. {\bf 93}, 050401 (2004).

\bibitem{huletBEC}
R. Hulet, KITP Conference on Quantum Gases, Santa Barbara, May 10 - 14, 2004.

%%%%%%%%% Feshbach
\bibitem{feshbacheite}
E. Tiesinga, B. J. Verhaar, and H. T. C. Stoof, Phys. Rev. A {\bf
47}, 4114 (1993).

\bibitem{feshbach}
S. Inouye {\it et al.}, %M. Andrews, J. Stenger, H.-J. Miesner, S. Stamper-Kurn, and W. Ketterle,
Nature {\bf 392}, 151 (1998).

%%%%%%%%%%% crossover expts %%%%%%%%%%%%%%%%%%%%%%%
\bibitem{markus1}
M. Bartenstein {\it et al.}, %, A. Altmeyer, S. Riedl, S. Jochim, C. Chin, J. Hecker Denschlag, and R. Grimm,
Phys. Rev. Lett. {\bf 92}, 120401 (2004).

\bibitem{jin2004}
C. A. Regal, M. Greiner, D. S. Jin, Phys. Rev. Lett. {\bf 92}, 040403 (2004).

\bibitem{mit2004}
M. W. Zwierlein {\it et al.}, %C. A. Stan, C. H. Schunck, S. M. F. Raupach, A. J. Kerman, and W.
%Ketterle,
Phys. Rev. Lett. {\bf 92}, 120403 (2004).

\bibitem{kinast2004}
J. Kinast {\it et al.}, Phys. Rev. Lett. {\bf 92}, 150402 (2004).

\bibitem{markus2}
M. Bartenstein {\it et al.}, %, A. Altmeyer, S. Riedl, S. Jochim, C. Chin, J. Hecker Denschlag, and R. Grimm,
Phys. Rev. Lett. {\bf 92}, 203201 (2004).

\bibitem{kinast2004b}
J. Kinast, A. Turlapov, and J. E. Thomas, Phys. Rev. A {\bf 70}, 051401(R)
(2004).

\bibitem{chin2004}
C. Chin {\it et al.}, Science {\bf 305}, 1128 (2004); published
online 22 July 2004 (10.1126/science.1100818).

%%%%%%%%%%%%%% resonance superfluidity

\bibitem{holland}
M. Holland, S.J.J.M.F. Kokkelmans, M. L. Chiofalo, and R. Walser, Phys. Rev.
Lett. {\bf 87}, 120406 (2001).

\bibitem{timmermans}
E. Timmermans, K. Furuya, P. W. Milonni, and A. K. Kerman, Phys. Lett. A {\bf
285}, 228 (2001).

%J.N. Milstein, S.J.J.M.F. Kokkelmans, and M. Holland, Phys. Rev. A
%{\bf 66}, 043604 (2002);

\bibitem{ohashi}
Y. Ohashi and A. Griffin, Phys. Rev. Lett. {\bf 89}, 130402 (2002).

\bibitem{stajic}
J. Stajic {\it et al.}, Phys. Rev. A {\bf 69}, 063610 (2004).


%%%%%%%%%% Li6 degeneracy

\bibitem{DFGRice}
A. G. Truscott {\it et al.}, Science {\bf 291}, 2572 (2003).

\bibitem{DFGENS}
F. Schreck {\it et al.}, Phys. Rev. Lett. 87, 080403 (2001).

\bibitem{DFGDuke}
S. R. Granade {\it et al.}, Phys. Rev. Lett. {\bf 88}, 120405 (2002).

\bibitem{DFGMIT}
Z. Hadzibabic {\it et al.}, Phys. Rev. Lett. {\bf 88}, 160401 (2002).

%%%%%%%%%%%%%%%%%%%%%%%%%%%%%%

\bibitem{houbiers}
M. Houbiers, H. T. C. Stoof, W. I. McAlexander, and R. G. Hulet, Phys. Rev. A
{\bf 57}, R1497 (1998).

\bibitem{huletnarrow}
K. E. Strecker, G. B. Partridge, and R. G. Hulet, Phys. Rev. Lett. {\bf 91},
080406 (2003).

\bibitem{thomasscience}
K. M. O'Hara {\it et al.}, Science {\bf 298}, 2179 (2002);
published online 7 Nov 2002 (10.1126/science.1079107).

\bibitem{bourdel}
T. Bourdel {\it et al.}, Phys. Rev. Lett. {\bf 91}, 020402 (2003).

\bibitem{ENSmolecules}
J. Cubizolles {\it et al.}, Phys. Rev. Lett. {\bf 91}, 240401 (2003).

\bibitem{IBKmolecules}
S. Jochim {\it et al.}, Phys. Rev. Lett. {\bf 91}, 240402 (2003).

\bibitem{markus3}
M. Bartenstein {\it et al.}, cond-mat/0408673.


%%%%%%%%% WEAKLY BOUND DIMERS etc.

\bibitem{petrov}
D. S. Petrov, C. Salomon, and G. V. Shlyapnikov, Phys. Rev. Lett. {\bf 93},
090404 (2004).

\bibitem{carr}
L. D. Carr, G. V. Shlyapnikov, and Y. Castin, Phys. Rev. Lett.
{\bf 92}, 150404 (2004).

\bibitem{chin03}
C. Chin and R. Grimm, Phys. Rev. A {\bf 69}, 033612 (2004).

\bibitem{french}
S. J. J. M. F. Kokkelmans, G. V. Shlyapnikov, and C. Salomon, Phys. Rev. A {\bf
69}, 031602 (2004).

%%%%%%%% UNIVERSALITY


\bibitem{heiselberg}
H. Heiselberg, Phys. Rev. A {\bf 63}, 043606 (2001).

\bibitem{universality}
T.-L. Ho, Phys. Rev. Lett. {\bf 92}, 090402 (2004).

\bibitem{thomasmechan}
M. E. Gehm \emph{et al.}, Phys. Rev. A {\bf 68}, 011401 (2003).

\bibitem{carlson}
J. Carlson, S.-Y. Chang, V.R. Pandharipande, and K.E. Schmidt, Phys. Rev. Lett.
{\bf 91}, 050401 (2003).

%%%%%%%%%%% theories

\bibitem{astrakharchik}
G. E. Astrakharchik, J. Boronat, J. Casulleras, and S. Giorgini, Phys. Rev.
Lett. {\bf 93}, 200404 (2004).

\bibitem{strinatiprofiles}
A. Perali, P. Pieri, and G. C. Strinati, Phys. Rev. Lett. {\bf 93} 100404
(2004). The comparison with our experiment assumed a Feshbach resonance center
at 850\,G. The actual position of 834\,G \cite{markus3} essentially changes the
best fit atom number to $1.95\times10^5$ instead of $2.3\times10^5$.

\bibitem{strinaticompare}
P. Pieri, L. Pisani, and G. C. Strinati, cond-mat/0410578.

%%%%%%%%%%%%%%%% collective excitations %%%%%%%%%%%%%%%%%%%

\bibitem{stringari96}
S. Stringari, Phys. Rev. Lett. {\bf 77}, 2360 (1996).

\bibitem{stringaribook}
L. Pitaevski and S. Stringari, {\it Bose-Einstein Condensation} (Clarendon
Press, Oxford, 2003), and refs.\ therein.

\bibitem{excite}
G. M. Bruun and C.W. Clark, Phys. Rev. Lett. {\bf 83}, 5415 (1999); G. M. Bruun
and B. R. Mottelson, Phys. Rev. Lett. {\bf 87}, 270403 (2001); A. Minguzzi and
M. P. Tosi, Phys. Rev. A {\bf 63}, 023609 (2001).

\bibitem{stringari04}
S. Stringari, Europhys. Lett. {\bf 65}, 749 (2004).

\bibitem{tosi04}
Hui Hu, A. Minguzzi, Xia-Ji Liu, and M. P. Tosi, Phys. Rev. Lett.
{\bf 93}, 190403 (2004).

%%%%%%%%%%%%%%%%%% rf spectroscopy %%%%%%%%%%%%%%%%%%%%%%%%%%%%%%

\bibitem{combescotcomment}
R. Combescot and X. Leyronas, Phys. Rev. Lett. {\bf 93}, 138901 (2004).

\bibitem{debbieRF}
C. Regal and D. Jin, Phys. Rev. Lett. {\bf 90}, 230404 (2003).

\bibitem{ketterleRF}
S. Gupta {\it et al.}, %, Z. Hadzibabic, M. W. Zwierlein, C. A. Stan, K. Dieckmann, C. H. Schunck, E.
%G. M. van Kempen, B. J. Verhaar, and W. Ketterle,
Science {\bf 300}, 1723 (2003); published online 8 May 2003
(10.1126/science.1085335).

\bibitem{jinnature1}
C. A. Regal, C. Ticknor, J. L. Bohn, and D. S. Jin, Nature {\bf 424}, 47
(2003).

%%%%%%%%%%%%%%%%%%%%%%%%%%%%%%%%%%%%%%%%%%%%%%%%%%%%%%%%%%%%%%%%%%%

\bibitem{zollergap1}
P. T\"{o}rm\"{a} and P. Zoller, Phys. Rev. Lett. {\bf 85}, 487 (2000).

\bibitem{paivi1}
J. Kinnunen, M. Rodriguez, and P. T\"orm\"a, Phys. Rev. Lett. {\bf 92}, 230403
(2004).

\bibitem{buechler}
H. P. B\"uchler, P. Zoller, W. Zwerger, Phys. Rev. Lett. {\bf 93}, 080401
(2004).

\bibitem{honew}
R. B. Diener and T.-L. Ho, cond-mat/0405174.

\bibitem{paivi2}
J. Kinnunen, M. Rodriguez, and P. T\"orm\"a, Science {\bf 305}, 1131 (2004);
published online 22 July 2004 (10.1126/science.1100782).

\bibitem{temperature}
Lacking a reliable method to directly determine the temperature $T$ of a deeply
degenerate, strongly interacting Fermi gas, we characterize the system by the
temperature $T'$ measured after an isentropic conversion into the BEC limit
\cite{chin2004}. Note that in general $T \le T'$ \cite{carr}.

\bibitem{chinjulienne}
C. Chin and P. Julienne, Phys. Rev. A, in press; cond-mat/0408254.

\bibitem{bulgac}
A. Bulgac, cond-mat/0309358.

\bibitem{strinati}
A. Perali, P. Pieri, L. Pisani, and G. C. Strinati, Phys. Rev. Lett. {\bf 92},
220404 (2004).

\bibitem{markusdr}
M. Bartenstein, doctoral thesis, University of Innsbruck (2005).


\end{thebibliography}
\end{document}